\documentclass{jfm}

\usepackage{graphicx}
\usepackage{natbib}
\usepackage{amsmath}
\usepackage{subfigure}
 
\usepackage{upmath}
\usepackage{amssymb}
\usepackage{amsbsy}

\usepackage{color}
\definecolor{red}{RGB}{255,0,0}

\newcommand\Rey{\mbox{\textit{Re}}}  
\newcommand\ReyD{\mbox{\textit{Re$_D$}}}  

\newsavebox{\astrutbox}
\sbox{\astrutbox}{\rule[-5pt]{0pt}{20pt}}

\title[Diffusive dynamics and stochastic models of turbulent axisymmetric wakes]{Diffusive dynamics and stochastic models of turbulent axisymmetric wakes}

\author[G. Rigas, A. S. Morgans, R. D. Brackston and J. F. Morrison]
{G. Rigas \thanks{Email address for correspondence: g.rigas@imperial.ac.uk} , A. S. Morgans, R. D. Brackston, J. F. Morrison}

\affiliation{Department of Aeronautics, Imperial College London, London SW7 2AZ, UK}

\pubyear{2010}
\volume{650}
\pagerange{119--126}
\date{?; revised ?; accepted ?. - To be entered by editorial office}

\begin{document}

\maketitle

\begin{abstract}
A modelling methodology to reproduce the experimental measurements of a turbulent flow under the presence of symmetry is presented. The flow is a three-dimensional wake generated by an axisymmetric body. We show that the dynamics of the turbulent wake-flow can be assimilated by a nonlinear two-dimensional Langevin equation, the deterministic part of which accounts for the broken symmetries which occur at the laminar and transitional regimes at low Reynolds numbers and the stochastic part of which accounts for the turbulent fluctuations. Comparison between theoretical and experimental results allows the extraction of the model parameters.
\end{abstract}


\section{Introduction}\label{sec:intro}

Turbulent flows are ubiquitous in natural phenomena and engineering applications \citep{pope2000turbulent} therefore a mathematically tractable  description of them is desirable for their prediction and control. At low Reynolds numbers, corresponding to laminar regimes, bifurcation theory has aided understanding of the dynamic behaviour of fluid flows in the presence of symmetry: under certain assumptions, the infinite-dimensional nonlinear fluid system governed by the Navier-Stokes equations can be reduced to a simple and finite-dimensional system close to the threshold of bifurcation \citep{crawford1991symmetry, Holmes2012}. However, departure from the laminar regimes and the critical bifurcating points renders the flow chaotic and finally turbulent, increasing the order of the system and the complexity for a mathematical description of it. 

Here we analyse the turbulent wake generated by an axisymmetric bluff body at high Reynolds numbers. This type of flow undergoes a steady bifurcation followed by an unsteady one  at low Reynolds numbers  \citep{Bohorquez2011, bury2012transitions} the dynamics of which can be accurately captured using simple weakly nonlinear models \citep{fabre2008bifurcations, meliga2009global} prior to the emergence of chaos and turbulent behaviour. During the transitional regime, continuous spatial and temporal symmetries are spontaneously broken. Although bifurcations break the symmetries en route to turbulence, fully developed turbulence is known to restore the possible symmetries in a statistical sense, at very high Reynolds numbers \citep{frisch1995turbulence}. 

In this paper we show that the dynamics of the turbulent wake can be approximated by a simple nonlinear Langevin equation, the deterministic part of which accounts for the broken symmetries which occur in the laminar and transitional regime, and the stochastic part for the turbulent fluctuations. 

The paper first presents an overview of axisymmetric wake-flows and models in the laminar (\S \ref{sec:laminar}) and turbulent (\S \ref{sec:turbulent}) regimes, before presenting the stochastic model in \S \ref{sec:model}. The model is validated using experimental data  (\S \ref{sec:exp}), and the results are presented in \S \ref{sec:results}.

\subsection{Laminar regime and modelling}\label{sec:laminar}

The laminar and linearly stable axisymmetric wake loses spatial rotational symmetry, $\mathcal{R}_\pi$, in the azimuthal direction (rotation of angle $\pi$ around any radial axis passing through the centre of the body) due to a supercritical pitchfork  bifurcation  \citep{fabre2008bifurcations, meliga2009global, Bohorquez2011, Bobinski2014}, the normal form of which reads
\begin{equation}
\label{eq:SL}
\newcommand{\x}{\mathbf{x}}
	\dot{\x}= \alpha \x + \lambda \x|\x|^2
\end{equation}
($\alpha, \lambda \in \mathbb{R},~\lambda<0$). Above the threshold of instability ($\alpha>0$), Eq.~\eqref{eq:SL} is associated with symmetry breaking since the steady state solutions are not invariant under the $\mathbf{x} \to -\mathbf{x}$ symmetry. This model can be directly obtained from the Navier-Stokes equations through a weakly nonlinear expansion  around the critical bifurcating point \citep{meliga2009global} and has been used extensively for the description of laminar flows undergoing supercritical bifurcations \citep{drazin2004hydrodynamic}. In the context of the weakly nonlinear analysis, $\mathbf{x}=\mathbf{x}(t)$ describes the complex amplitude evolution of the bifurcated state (global mode).

Increasing the Reynolds number, a subsequent Hopf bifurcation is responsible for the loss of continuous temporal symmetry and the emergence of periodic shedding of vortices before the turbulent axisymmetric wake reaches a chaotic regime \citep{bury2012transitions}.


\subsection{Turbulent regime}\label{sec:turbulent}

At high Reynolds numbers, a large number of studies have shown that the bifurcating states observed in the laminar wakes persist and manifest as large-scale structures. For the axisymmetric wake, it was recently shown that the  steady bifurcated state (responsible for the loss of rotational symmetry) and the unsteady bifurcated state (responsible for the vortex shedding)  persist at high Reynolds numbers and fully turbulent regimes \citep{Rigas2014JFMr, grandemange2014statistical}. Similar behaviour has been observed in wakes with other types of symmetries (see for example the rectilinear three-dimensional Ahmed body \citep{GrandemangePRE, grandemange2013turbulent}).

Here we extend the laminar weakly nonlinear modelling approach to the fully turbulent regime, and we show that the derived model captures the  dynamic evolution of the turbulent three-dimensional wake. This is achieved by modelling the effect of the turbulent fluctuations acting on the deterministic dynamics of the system, reminiscent of the transitional instabilities observed in the laminar regimes, as stochastic forcing. The stochastic modelling approach of the turbulent dynamics has been successfully applied to various turbulent flows exhibiting global and large-scale dynamics including the turbulent   Rayleigh-B{\'e}nard convection \citep{Sreenivasan2002Mean, brown2007large} and the turbulent von K{\'a}rm{\'a}n swirling flow \citep{de2007slow}. Hereafter we demonstrate our modelling approach considering the dynamics associated with the first steady bifurcation given by Eq.~\eqref{eq:SL}.

\section{The stochastic model}\label{sec:model}

\begin{figure}
\centering
\includegraphics[width=0.8\textwidth]{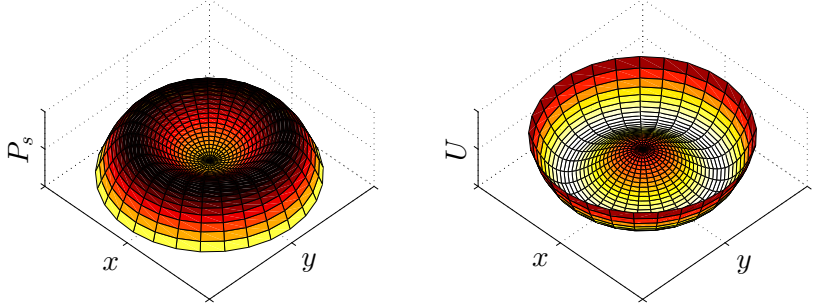}
\caption{\label{fig:potential} Stationary probability density function (left) and potential (right) of the nonlinear Langevin model.}
\end{figure}


We consider the deterministic system given by  Eq.~\eqref{eq:SL} and apply a stochastic modelling approach for the turbulent background fluctuations. In Cartesian coordinates, $\mathbf{x}=(x,y)$,  the deterministic system with independent additive white noise becomes
\begin{eqnarray}
\label{eq:Lagevin_cart}
	\dot{x} &=& \alpha x + \lambda x(x^2+y^2) + \sigma \xi_x(t), \notag \\
	\dot{y} &=& \alpha y + \lambda y(x^2+y^2) + \sigma \xi_y(t),
\end{eqnarray}
where $\mathbf{\xi}$ accounts for the random forcing of turbulence and $\sigma^2$ the variance of it. Here, $\sigma \equiv \sigma_x = \sigma_y$ and the stochastic process in Eq.~\eqref{eq:Lagevin_cart} is rotationally symmetric. We transform the system from $(x,y)\to (r,\phi)$, where $r$ is the radial distance from the centre (amplitude) and $\phi$ the angle (phase), using the Ito interpretation  \citep{gardiner1985stochastic}. In polar variables, the Langevin system given by Eq.~\eqref{eq:Lagevin_cart} becomes
\begin{eqnarray}
	\label{eq:Lagevin_polar}
	\dot{r} 	=  \alpha r + \lambda r^3  + \frac{\sigma^2}{2r} + \sigma \xi_r, \quad
	\dot{\phi}  =  \frac{\sigma}{r} \xi_\phi,
\end{eqnarray}
where $r=\sqrt{x^2+y^2}$ and $\phi = \tan^{-1}(y/x)$. Notice that in polar variables the radial component is independent of the angular position.

The stationary probability density  function (PDF) of the above system can be found from the steady state Fokker--Planck equation and is given by
\begin{equation}
P_s(r,\phi) = C \exp \left[ - \frac{U(r,\phi)}{K} \right],
\label{eq:SFP_r} 
\end{equation}
where $C$ is a normalisation constant, $K=\sigma^2/2$ is the noise intensity (diffusivity) and $U$ the potential. The potential $U$ is
\begin{equation}
U(r,\phi) =  -\left[ \frac{\alpha r^2}{2} +\frac{\lambda r^4}{4} + K \ln r  \right].
\end{equation}
The stationary PDF for the angle is uniform, $ \int_{0}^{\infty}P_s(r,\phi) d r=\frac{1}{2\pi}$. In the case of a rotationally  symmetric experimental setup and inflow conditions, the drifts and diffusivities are independent of $\phi$. The stationary PDF, Fig.~\ref{fig:potential}, peaks around the minimum of the Mexican-hat shaped potential.

\section{Experimental set-up}\label{sec:exp}
\begin{figure}    
\begin{center}
 \includegraphics[width=0.95\textwidth]{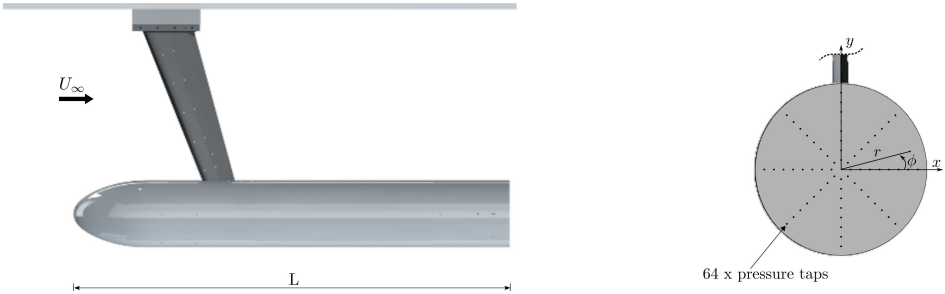}
 \caption{Schematic representation of the model. (a) Side view; (b) rear view. Pressure measurements are taken on the base of the body from 64 static pressure tappings, based on which the Centre-of-Pressure location is calculated. }
  \label{FIG1}
\end{center}
\end{figure}

The turbulent wake is generated by an axisymmetric bluff-body with a blunt trailing edge, a schematic of which is shown in Fig.~\ref{FIG1}. The body diameter $D$, is 196.5 mm and the length-to-diameter ratio $L/D$, is 6.48.  More details for the experimental setup and procedure can be found in \cite{Rigas2014JFMr} and \cite{oxlade2015high}. Experiments were performed at a constant free stream velocity, $U_\infty=15$ m/s. The Reynolds numbers based on diameter and boundary layer momentum thickness at separation are $\ReyD=1.88 \times 10^5$ and $\Rey_\theta = 2050$ respectively.  The boundary layer separation and the ensuing wake are turbulent.
 
Pressure measurements are taken on the base of the body from 64 static taps equally spaced on a uniform polar grid with $\delta r=0.056D$ and $\delta \phi=45^{\circ}$ in the radial and azimuthal directions, respectively. The sampling frequency was 225 Hz. A total of 19,200 s of data was acquired over sixteen independent experiments, providing approximately 2000 independent measurements with a 95\% uncertainty of approximately 0.45\% and 1\% in time average and rms pressure respectively.

\section{Results}\label{sec:results}

Let $\mathbf{x}$ of Eq.~\eqref{eq:SL} represent the Centre-of-Pressure (CoP) coordinates. The CoP is used as a macroscopic variable quantifying the symmetry of the turbulent wake. Here it is calculated from the space-averaged pressure, defined on the Cartesian coordinate system of the base, $\mathbf{x}'=(x',y')$, and non-dimensionalised with the body diameter, as
$$
\mathbf{x}(t) = \frac{1} {\int \! p(t) dA}   \int_A p(t) \mathbf{x}' dA,
$$
where $A$ is the area of the base of the body. A zero CoP value corresponds to an $\mathcal{R}_\pi$-symmetric flow, whereas departure from this value to an increased asymmetry of the flow \citep{Rigas2014JFMr}.

\begin{figure}
\centering
 \includegraphics[scale=0.5]{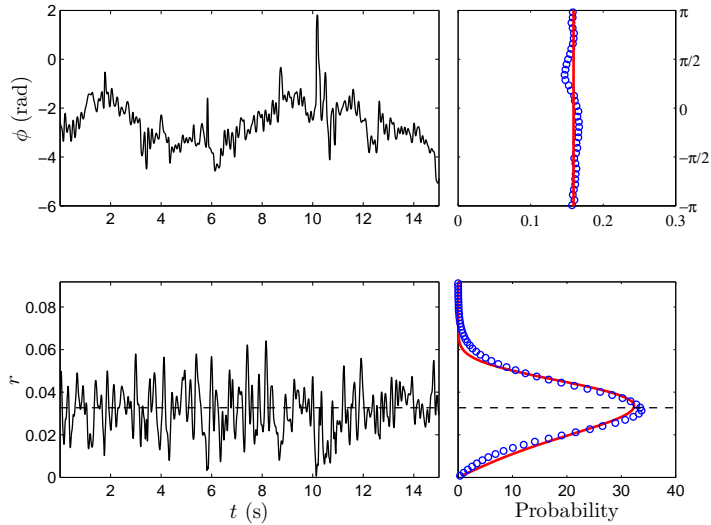}
 \caption{\label{FIG2} A time series  and probability density of the Centre-of-Pressure: angular (upper) and radial (lower) position. Symbols in the probability density (right) correspond to experimental data and solid lines to theoretical model predictions. Dashed line: mean radial value indicating the broken $\mathcal{R}_\pi$ symmetry.}
 \end{figure}
 
The temporal evolution and PDF of the radial and azimuthal location of the CoP for the highly turbulent regime  are shown in Fig.~\ref{FIG2}.  Highly erratic motion with time is observed in both components. Statistically, the wake spends most of the time in a non-zero radial location, indicating broken $\mathcal{R}_\pi$ symmetry. Due to the uniform PDF in the azimuthal location, rotational symmetry is recovered in the long time average. For the laminar regime and after the first bifurcation, which is described by Eq.~\eqref{eq:SL}, the stable equilibrium corresponds to a non-zero CoP, the angle of which is determined from the initial conditions and is unique. During the turbulent regime, the laminar stable equilibrium point explores a continuum of states around its mean radial value and uniformly distributed in the angular direction: hence, the wake explores an infinite number of metastable states restoring the lost $\mathcal{R}_\pi$ symmetry. 

 \begin{figure}
 \centering
 \includegraphics[scale=0.4]{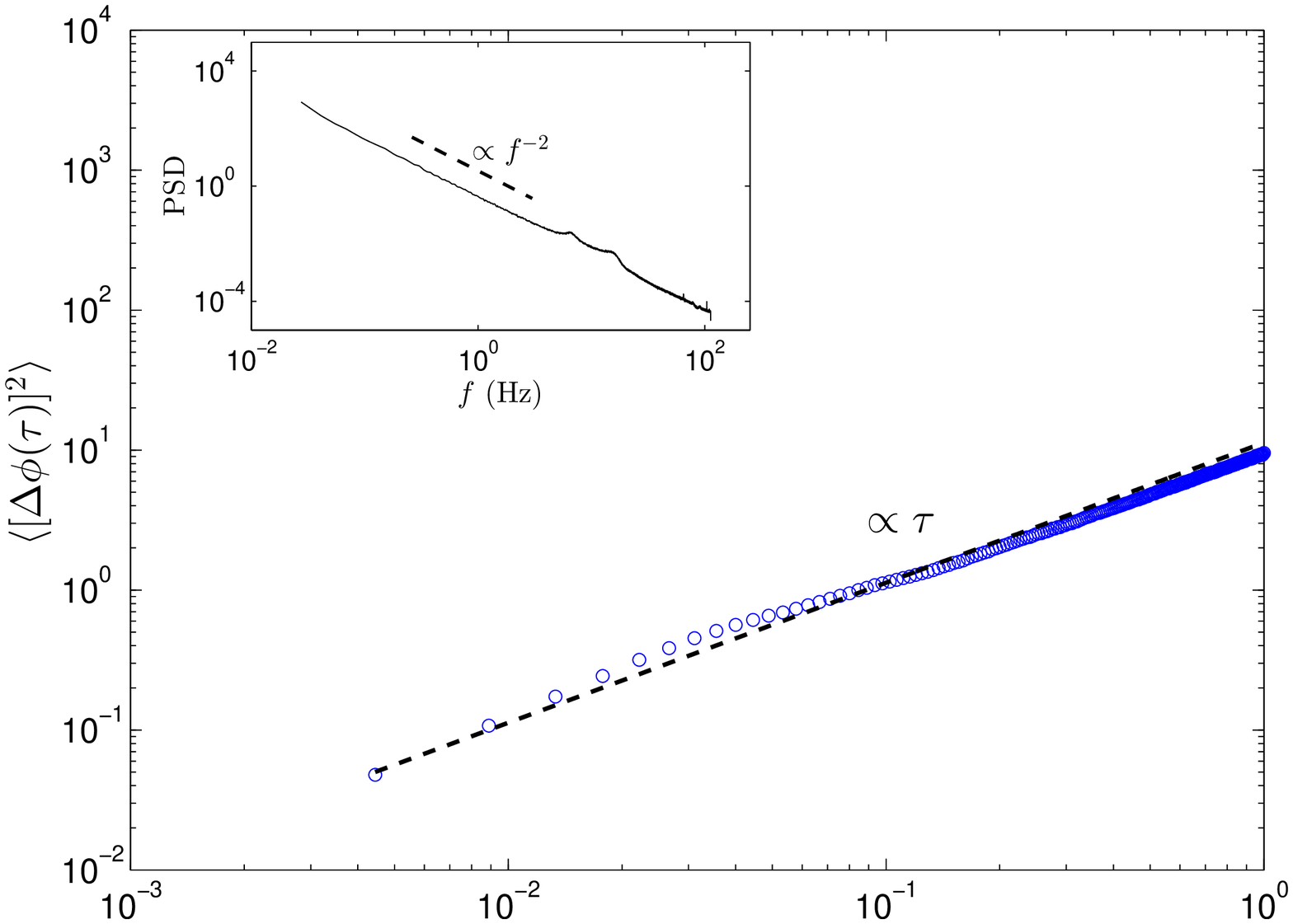}\\
  \includegraphics[scale=0.4]{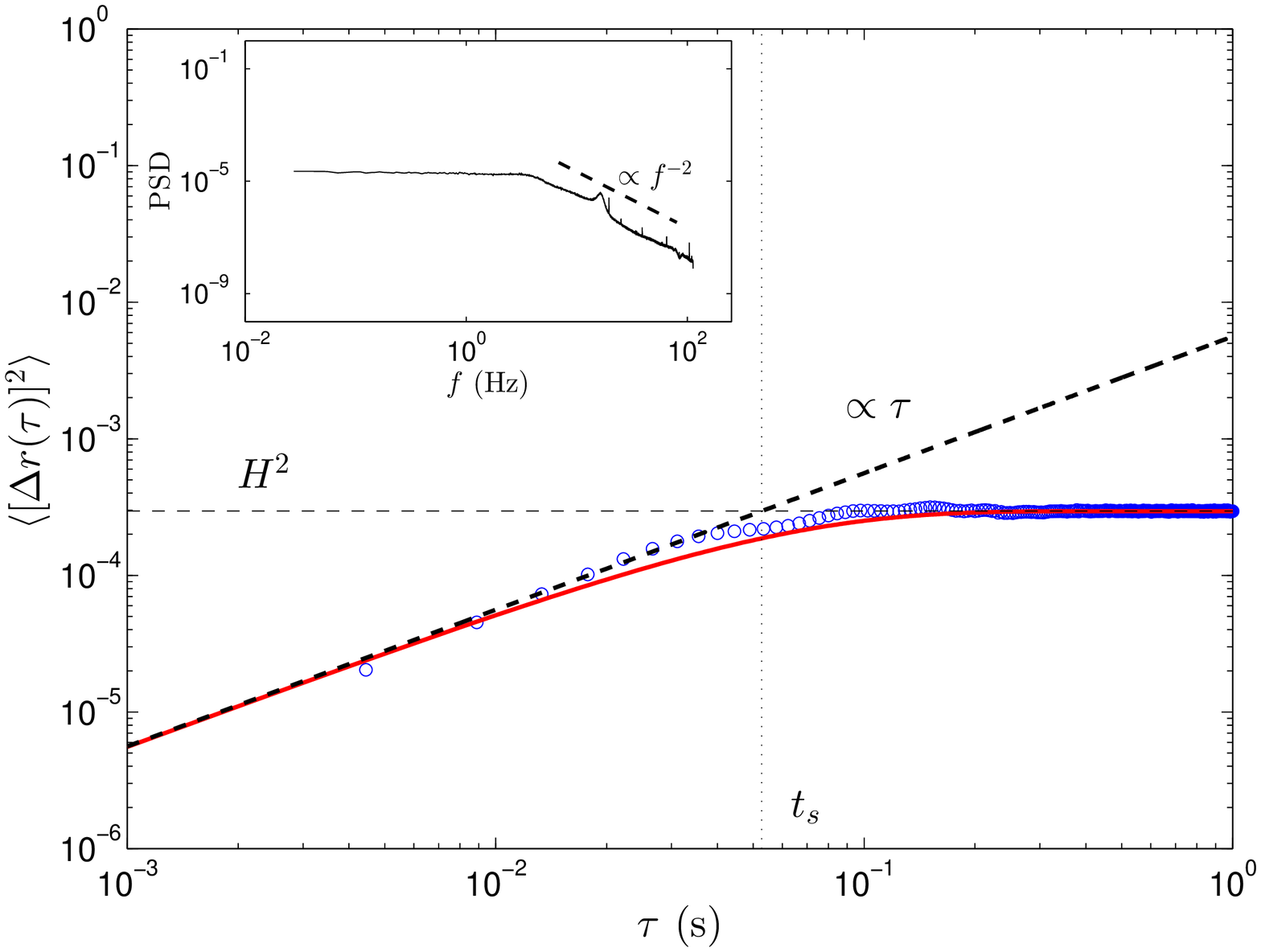}
 \caption{\label{fig:msd} Mean square displacement as a function of the time interval $\tau$. (a) Angular and (b) radial components of the CoP location. Symbols: experimental data. Solid line: model. Thick dashed lines: power-law fits. Insets show power spectral density of the experimental CoP components. An exponent of -2, consistent with free diffusive motion, is obtained for the azimuthal and radial components. At low frequencies, the PSD of the radial components is constant due to the spatial confinement imposed by the potential well, in accordance with the MSD results.}
 \end{figure}

Insight into the random dynamics of the turbulent wake is provided from the calculation of the time-averaged mean-square displacement (MSD), defined as $\left<[\Delta \mathbf{x}(\tau)]^2 \right> = \left<  ( \mathbf{x}(t+\tau) - \mathbf{x}(t) )^2  \right>$. The MSD of the angular and radial components is plotted in  Fig.~\ref{fig:msd}  for different sampling times. In the azimuthal direction, the MSD increases linearly with time, $\left<[\Delta \phi(\tau)]^2 \right>\propto \tau$, consistent with free diffusive motion \citep{einstein1905molekularkinetischen}. In the radial direction, the linear relation holds only for short time scales below a threshold $t_s$, $\left<[\Delta r(\tau)]^2 \right>\propto \tau, ~ \tau<t_s$, and reaches a saturation plateau at larger time scales, $\lim_{\tau\to\infty} \left<[\Delta r(\tau)]^2 \right> = H^2$. 

The above results for the CoP are consistent with the predictions of the model given by Eq.~\eqref{eq:Lagevin_polar}.    The coefficients of the Langevin equation $\alpha$, $\lambda$ and $K$, are obtained from the experimental MSD and PDF and are given in Table~\ref{tab:table1}.  The slope of the radial MSD relation is directly correlated with the diffusion coefficient $K$, which is obtained through linear fitting, $\langle [\Delta r(\tau)]^2\rangle = 2K\tau,  ~ \tau<t_s$.  Knowing the diffusion coefficient $K$, the model coefficients $\alpha$, $\lambda$ are uniquely defined from Eq.~\eqref{eq:SFP_r} and here are obtained through least-square fitting. 

In Fig.~\ref{FIG2}, we compare the  measured probability density function with the one predicted by the model from Eq.~\eqref{eq:SFP_r}.  In Fig.~\ref{fig:msd} we compare the MSD of the experimental results with the numerically calculated MSD from the model. Direct numerical integration of  Eq.~\eqref{eq:Lagevin_polar} was performed using an Euler-Maruyama scheme. We notice that the dynamics of the CoP can be described over all the time scales from the Langevin model.

\begin{table}
\begin{center}
\def~{\hphantom{0}}
\setlength{\tabcolsep}{12pt}
\begin{tabular}{lcccc}
$\alpha$ 	& $\lambda$ 	& $K=\sigma^2/2$ \\
3.81 		& -5604 		& 0.0028 \\
\end{tabular}
\caption{\label{tab:table1}%
Coefficient values of the Langevin model obtained experimentally at $\ReyD=1.88 \times 10^5$. The units of the coefficients are s$^{-1}$ since lengths are non-dimensionalised with the body diameter.}
\end{center}
\end{table}

A qualitative explanation of the above results is provided by the potential well shown in Fig.~\ref{fig:potential}. The turbulent wake, the state of which is quantified by the CoP, meanders in the Mexican-hat-shaped potential and explores an infinite number of states through a random walk (diffusive motion). Specifically, in the azimuthal direction, which determines the orientation of the wake,  it can explore freely any azimuthal location resulting in unbounded reorientations. In the radial direction, the motion is restricted due to spatial constraints  (confinement imposed by the potential well) resulting in a constant MSD at large timescales.
 
The diffusive dynamics of the turbulent wake are also depicted in the power spectral density of the CoP, plotted as inset in Fig.~\ref{fig:msd}. The spectral density is closely related to the MSD: in general, for a power law behaviour of the MSD, $\left<[\Delta \mathbf{x}(\tau)]^2 \right>\propto \tau^\kappa$, the asymptotic form of the power spectral density is $\Phi(f) \propto f^{-(1+\kappa)}$.  In accordance with the linear increase of the MSD with time ($\kappa=1$), an exponent of -2 is observed in the spectrum of the angular component consistent with Brownian motion. A similar decay is observed for the radial component when $f>1/t_s$. However, at low frequencies corresponding to ${f \to 0}$, or equivalently large timescales, it reaches a plateau and levels off, in accordance with the MSD measurements for $\tau\to\infty$. 
 
\begin{figure}
\centering
 \includegraphics[scale=0.6]{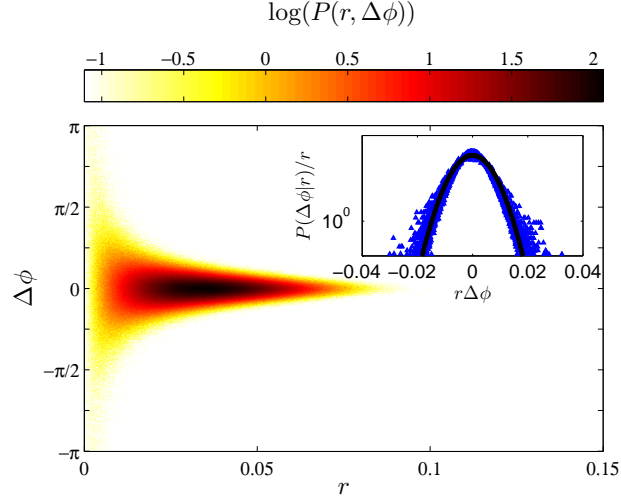}
 \caption{\label{fig:jointPDF} Reorientation characteristics: Joint PDF of angular variation and radial position of the CoP as measured from the experiment. Inset shows normalised conditional probabilities: data (symbols) collapse on a Gaussian distribution (line), as described by the model.}
 \end{figure}

The dynamics associated with the radial and angular motion of the CoP have been analysed independently so far.  Now, we analyse the coupled dynamics predicted by the model. Specifically the coupling arises as an inverse relationship between $\dot{\phi}=\Delta\phi/\Delta t$ and $r$ for the angular component, as described by Eq.~\eqref{eq:Lagevin_polar}. The model suggests that the conditional PDF of  $\dot{\phi}$ for a given $r$ follows a Gaussian distribution with variance inversely proportional to $r^2$, that is
\begin{equation}
	P(\dot{\phi}|r) = \mathcal{N}\left(0, \frac{\sigma^2}{r^2}\right) = \frac{r}{\sigma\sqrt{2\pi}}\, e^{- \frac{1}{2} \left( \frac{r\dot{\phi}}{ \sigma} \right)^2}.
	\label{eq:Joint_reor1}
\end{equation}
Indications of the inverse relationship can be found in the time series of the CoP in Fig.~\ref{FIG2}. One observes that for small values of the radial component, abrupt changes of the azimuthal orientation occur.  The inverse relation is further validated from the joint PDF of the angular variation and radial location, $p( r, \Delta \phi)$, shown in Fig.~\ref{fig:jointPDF}, where large reorientations are more probable at small radii. The conditional PDFs of the reorientations $\Delta \phi$, $p(\Delta \phi | r)=p(r, \Delta \phi)/P(r)$ collapse to a zero-mean Gaussian distribution with variance $\sigma^2$, when scaled by $1/r$ and plotted against $r\dot{\phi}$.  In the limit of small radii values, $r \to 0$, angular rotations with an almost uniform PDF are observed in the joint PDF. This situation of small $r$ corresponds to  large variance and therefore a very broad Gaussian that, over the finite measurable window of $-\pi$ to $\pi$, will appear uniform. This is analogous to the cessation events observed by \citet{brown2005reorientation} during the reorientation of the large-scales of Rayleigh-B{\'e}nard convection and can be understood without specifying arbitrary thresholds for a small $r$.
 
\section{Conclusions}\label{sec:conclusions}
The physical picture that can be drawn for the turbulent axisymmetric wake based on the above is as follows. The laminar large scale structures, associated with spatially broken symmetries, persist at high Reynolds numbers. In the turbulent regime, these structures undergo diffusive motion (random walk) in a two-dimensional Mexican-hat-shaped potential well (see Fig.~\ref{fig:potential}) restoring statistically the broken symmetries.  In this paper we have shown that this behaviour can be captured by a simple nonlinear model consisting of two coupled stochastic differential equations. The deterministic part of the model accounts for the broken symmetries observed in the laminar regime, and the stochastic part models in a phenomenological sense the turbulent fluctuations acting on the large scale structures. The unsteady bifurcation, responsible for the vortex shedding (limit cycle), has not been considered  in the present analysis. This can be done by adding an extra equation accounting for the temporal broken symmetries due to a Hopf bifurcation \citep{fabre2008bifurcations, meliga2009global}; experimentally, the quasi-periodic nature of the vortex shedding appears in the PSD of the CoP as peak at $\sim16$ Hz corresponding to a Strouhal number of $\sim0.2$ based on $D$ and $U_{\infty}$.

The diffusive dynamics of the large scale structures presented here and in  \cite{Rigas2014JFMr} show close similarities with the diffusive dynamics of the large-scale circulation observed in the turbulent Rayleigh-B{\'e}nard convection \citep{brown2005reorientation, brown2006rotations} and the turbulent von K{\'a}rm{\'a}n swirling flow \citep{de2007slow, berhanu2009bistability, petrelis2009simple}. However, this similarity is not surprising if the symmetries of the problem are accounted for: spatial rotational symmetry which breaks en route to turbulence.

In summary, we have shown that turbulent dynamics can be described by deterministic equations derived from symmetry arguments with stochastic forcing terms that give rise to turbulent dynamics. The modelling process we have presented here should be applicable to other turbulent flows, provided that their specific symmetries are taken into account.   \\

This work was supported by the Engineering and Physical Sciences Research Council (grant number EP/I005684).

\bibliographystyle{jfm}
\bibliography{jfmr_2015}

\end{document}